\documentclass[aps,pra,twocolumn,showpacs,groupedaddress]{revtex4-1}

\usepackage{graphicx,subfigure,color,verbatim,amsmath}

\newcommand{\sub}[1]{_{\mbox{\scriptsize #1}}}
\newcommand{\be}{\begin{equation}}
\newcommand{\ee}{\end{equation}}

\begin{document}

\title{Effect of trap anharmonicity on a free-oscillation atom interferometer}

\author{R. H. Leonard}
\author{C. A. Sackett}
\email[]{sackett@virginia.edu}
\affiliation{Department of Physics, University of Virginia, Charlottesville, Virginia, 22904, USA}

\date{\today}

\begin{abstract}
A free-oscillation interferometer uses atoms confined in a harmonic
trap. Bragg scattering from an off-resonant laser is used to split
an atomic wave function into two separated packets. 
After one or more oscillations in the trap, the
wave packets are recombined by a second application of the Bragg laser
to close the interferometer.
Anharmonicity in the trap potential
can lead to a phase shift in the interferometer output.
In this paper, analytical expressions for the anharmonic phase are derived
at leading order for perturbations of arbitrary power in the 
position coordinate.
The phase generally depends on the initial position and velocity of the 
atom, which are themselves typically uncertain. 
This leads to degradation in the interferometer performance,
and can be expected to limit the use of a cm-scale device
to interaction times of about 0.1~s. Methods to improve performance
are discussed.
\end{abstract}


\maketitle

\section{Introduction}

Atom interferometry is a sensitive tool for metrology and
probes of fundamental constants \cite{Berman97,Cronin09}.
Traditionally it uses atoms in free
space, but many groups have investigated interferometry of trapped atoms,
both thermal and Bose-condensed
\cite{Shin04,Wang05,Wu05,Shumm05,Garcia06,Arnold06,Horikoshi07,Sapiro09,Stickney09,Baumgartner10}. 
The use of confined atoms has several advantages, including the
abilities to maintain high density for interaction studies
and to impose complex atomic trajectories via the trapping potential.
Another significant 
benefit of confinement is that the atoms do not fall under gravity,
so measurements can be extended to long times without requiring a long
drop distance. 

\begin{figure}
\includegraphics[width=3.5in]{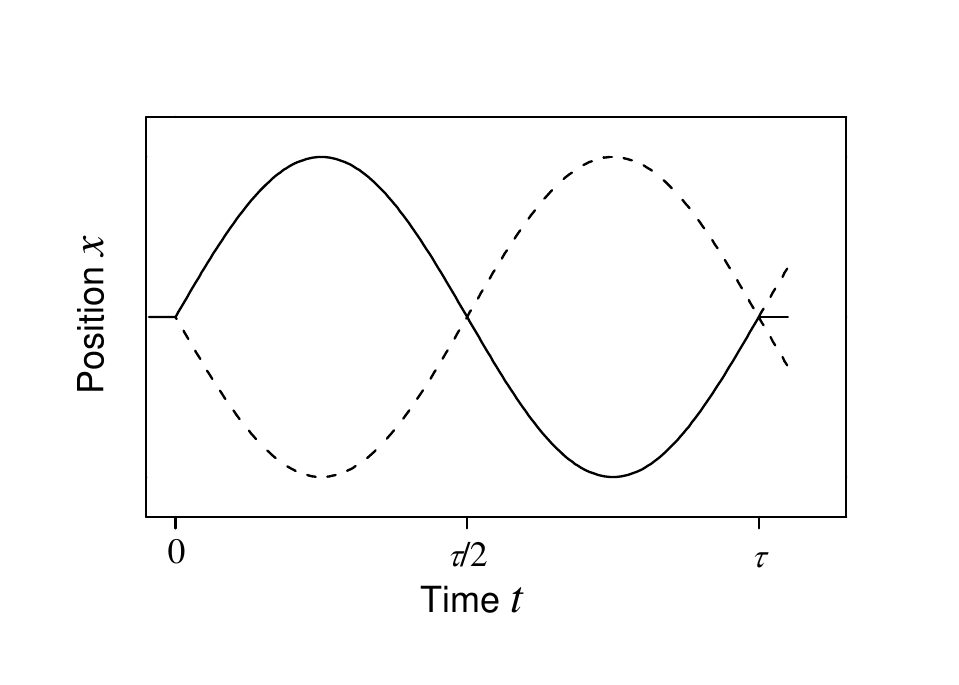}
\caption{
\label{trajectories}
Atomic trajectories $x(t)$ in a free-oscillation interferometer. 
Cold atoms are confined in a harmonic potential with period $\tau$. At time
$t = 0$ a laser pulse splits the atomic wave function into two packets
(solid and dashed curves)
moving in opposite directions. After completing either a half or a full
oscillation in the trap, the packets are recombined. (A full oscillation
is shown.) The fraction
of atoms that returns to rest depends on the phase difference
developed between the two packets.
}
\end{figure}

One interferometer configuration that has been of recent interest is
the ``free-oscillation'' interferometer (Fig.~\ref{trajectories}), 
in which atoms are confined
in a harmonic trap \cite{Horikoshi07,Burke08,Stickney09,Segal10,Kafle11}. 
An off-resonant laser pulse is applied
to the atoms and induces momentum kicks via Bragg scattering.
Typically, the laser is a standing wave tailored to produce
two wave packets with momentum kicks $\pm2\hbar k$ for light with
wavenumber $k$. The two packets
are allowed to separate and complete a half or full oscillation in the 
trap, at which time the laser pulse is applied again. The fraction of
atoms brought back to rest after the second pulse depends on the
phase difference between the packets at the time of recombination,
making the device an interferometer. 

The free-oscillation interferometer
can be compared with the Michelson configuration
\cite{Wang05,Wu05,Garcia06,Stickney08}, in which the atoms are
held in a potential that is nearly flat in one direction. The wave
packets are separated in that direction and one or more Bragg
reflection pulses are used to reverse the atomic motion and 
bring the packets back together. The free-oscillation interferometer
offers several advantages over the Michelson interferometer,
chiefly stemming from the difficulty of achieving a perfectly
flat potential \cite{Olshanii05,Burke08,Kafle11}.

However, it is not easy to achieve a perfectly harmonic potential
either. Anharmonicity is likely to be an important limit to
the performance of a free-oscillation interferometer because it causes a phase
difference between the two arms that depends on the position
and velocity of the initial unsplit packet. Thermal atoms will
have a large spread in position and velocity, so anharmonicity
can be expected to severely limit the usability of a free-oscillation
interferometer with non-condensate atoms. Even a Bose-Einstein condensate
will have some variation in the initial conditions due to 
the uncertainty principle, along with
extraneous motional excitations from technical effects.

In this paper, we calculate the phase shift of a free-oscillation
interferometer that is induced by anharmonicity.  
We will ignore atomic interactions here,
although they can in fact be important \cite{Kafle11}.
In Section \ref{calc} we calculate the anharmonic phase shift.
In Section \ref{impact} we estimate typical magnitudes of the anharmonicity
and suggest some techniques for mitigation of the effect.
Finally, Section \ref{conclusion} presents concluding remarks.

\section{Phase Calculation}
\label{calc}

\subsection{Perturbative Approach}

With reference to the atomic trajectories shown 
in Fig.~\ref{trajectories},  
we take the atoms to travel along
the $x$ direction. We assume this is a principle axis of the trapping
potential, and also that anharmonicity does not introduce
significant coupling between $x$ and the other directions. 
In this case, the relevant motion is
governed by the one-dimensional Hamiltonian 
\be
H = \frac{p^2}{2m} + \frac{1}{2} m\omega_0^2 x^2 + m f(x),
\label{Hamiltonian}
\ee
where 
$m$ is the atomic mass, 
$p = m\dot{x}$ is the momentum,
$\omega_0$ is the unperturbed oscillation frequency, 
and $m f(x)$ is a perturbing potential, assumed to be small.

The phase accumulated by a packet is determined using the 
classical action \cite{Storey94,Berman97},
\be
\phi(t)-\phi(0) = \frac{1}{\hbar}\int_0^t \left(T - V\right)\, dt
\label{action}
\ee
where the kinetic and potential energies $T$ and $V$ are calculated
for the classical trajectory $x\sub{class}(t)$. To be precise, this gives the
phase of the packet wave function $\psi(x,t)$ at the packet center
$\langle x \rangle = x\sub{class}$. We here ignore the possible impact
of phase gradients across the packet, which require the 
inclusion of atomic interactions
to correctly analyze \cite{Olshanii05,Burke08,Kafle11}.

The classical trajectory is determined using the equation of motion
\be
\ddot{x} + \omega_0^2x = -\frac{df}{dx}.
\label{motion}
\ee
Perturbative techniques for the solution of \eqref{motion} are
well known \cite{Landau76,Barger95}.
We write $x = x_0 + x_1$,
where the subscript indicates the perturbative order. We take 
$x_0 = A\cos(\omega t-\theta)$, where $\omega = \omega_0 + \omega_1$
includes an amplitude-dependent frequency shift $\omega_1$. 
Substituting these
expressions into \eqref{motion} and collecting the first-order
terms yields
\begin{align}
\ddot{x_1} + \omega_0^2 x_1 
= 2\omega_0\omega_1 x_0 - \left.\frac{df}{dx}\right|_{x_0}
\label{motion1}
\end{align}

The frequency shift $\omega_1$ is determined by requiring
that the right-hand side of \eqref{motion1} contains no terms oscillating
at $\omega_0$, since these would drive unbounded excitation of $x_1$.
To first-order, this yields
\be
\omega_1 = \frac{1}{2\pi \omega_0 A}\int_0^{2\pi} \cos u 
\left.\frac{df}{dx}\right|_{x = A\cos u}\, du
\label{w1}
\ee

For the interferometer phase shift, 
we consider the case where atoms make one full oscillation
in the potential
before recombination, as shown in Fig.~\ref{trajectories}.
The phase developed by a wave packet during
this motion is, from \eqref{action},
\be
\phi = \frac{m}{\hbar}\int_0^\tau 
\left(\frac{\dot{x}^2}{2} - \frac{\omega_0^2x^2}{2} - f(x) 
\right)\,dt.
\label{action2}
\ee
where $\tau$ is the period.
Using $x = x_0+x_1$ and neglecting terms higher than first-order yields
\be
\phi_1 = \frac{m}{\hbar}\int_0^\tau
\left[\frac{1}{2}\left(\dot{x}_0^2 - \omega_0^2 x_0^2\right)
-f(x_0) \right]\, dt,
\label{phase1}
\ee
The terms involving $x_0 x_1$ vanish upon integration because
$x_0$ and $x_1$ are, by construction, orthogonal on this interval.
To first-order, the derivative term in \eqref{phase1} can be expressed as
\be
\dot{x}_0^2 = (\omega_0^2 + 2\omega_0\omega_1)A^2\sin^2(\omega t-\theta).
\ee
Using this to evaluating the integral leaves a perturbation phase
\be
\phi_1 = \frac{\pi m}{\hbar} A^2 \omega_1 
- \frac{m}{\hbar} \int_0^\tau f(x_0)\,dt.
\label{phase2}
\ee

Two cases can now be considered. The first supposes that $f(x)$ varies
over a length scale that is small compared to the amplitude of the atomic motion
$A$. This might be caused by speckle in a laser trap or roughness
in the conductors of a magnetic chip trap. In this case, $\omega_1$ will
be small because $df/dx$ in Eq.~\eqref{w1} will rapidly
oscillate. In this limit,
the phase shift will be dominated by the second term in 
\eqref{phase2}, which is simply the integral of the perturbing potential:
\be
\phi_1 \rightarrow - \frac{m}{\hbar} \int_0^\tau f(x_0)\,dt.
\label{phase_s}
\ee
Further analysis would require knowing $f(x)$, which will in general
be specific to a particular apparatus. 

The second case supposes that $f(x)$ is slowly varying compared to the
atomic amplitude. This might result when the confinement potential is
created by a distant magnetic or optical element. 
Here it is reasonable
to Taylor expand $f(x)$ as a power series
\be
f(x) \rightarrow \sum_{n=3}^{\infty} f_n \left(\frac{x}{R}\right)^n
\label{power}
\ee
where $R$ is a characteristic length scale, which can
typically be taken as the distance to the trapping element.
If $x \ll R$, then the dominant term will be the lowest power of $n$
for which $f_n$ is non-vanishing. 
In a symmetric trap, $f_n$ will
be suppressed for all odd $n$, and careful design may result
in the suppression of $f_n$ for one or more even $n$ as well.
In general, however,
$\omega_1$ will be non-zero and both terms of Eq.~\eqref{phase2}
will contribute. 
This case corresponds to the conventional
anharmonic oscillator, and it is the main focus of the present paper.
We continue its analysis in Section \ref{firstorder} below.

In the intermediate case, where the length scale of $f$ is comparable
to $A$, the effect on the phase is complicated. If the detailed
form of $f(x)$ is known, then Eqs.~\eqref{w1} and \eqref{phase2} can be used
to numerically estimate the phase shift. However, direct numerical computation
of Eq.~\eqref{action} would require similar effort and give greater
accuracy.

\subsection{Anharmonic Trap: First Order}
\label{firstorder}

As explained above, we here consider the case of a power-law perturbation
which we will express as
\be
f(x) = \frac{1}{n} \lambda_n x^n
\label{lambda}
\ee
If $n$ is odd, then the result is simple: 
$\omega_1$ in Eq.~\eqref{w1} and $\phi$ in Eq.~\eqref{phase2} are both
zero by symmetry. There is thus no first-order effect on the interferometer,
but we will consider the second order effect in Section \ref{secondorder}.

If $n$ is even, then the frequency shift can be expressed as
\be
\omega_1 = \frac{\lambda_n}{2\pi} A^{n-2} 
\int_0^\tau \cos^n(\omega t-\theta)\,dt 
= \frac{\lambda_n}{\omega_0}A^{n-2} h_n
\label{w1_even}
\ee
where 
\be
h_n = \frac{1}{2\pi}\int_0^{2\pi} \cos^n u \,du = \frac{n!}{(n!!)^2}. 
\label{hn}
\ee
Using this in Eq.~\eqref{phase2} yields a trajectory phase
\be
\phi_1^{(n)} = \left(1-\frac{2}{n}\right)\frac{\pi m}{\hbar\omega_0} A^n \lambda
h_n
\qquad (n~\text{even}).
\label{evenphase}
\ee

Equation~\eqref{evenphase}
gives the first-order phase acquired by a wave packet during
one complete orbit through the confining potential.
However, in the actual interferometer, neither packet makes a 
perfectly complete orbit.
Suppose the wave packet starts at mean position $x_a$ and 
mean velocity $v_a$. It will be convenient to work with the
scaled velocity $u_a \equiv v_a/\omega_0$. 
The initial packet is split into two, with one packet acquiring a
velocity impulse $+u_0 = 2\hbar k/(m\omega_0)$ and the other $-u_0$.
The 
two packets therefore have motional amplitudes $A_\pm$ given
by $A_\pm^2 = x_a^2 + u_a^2 + u_0^2 \pm 2u_au_0$. Because these
amplitudes are different, the packets will experience different
motional frequencies $\omega_\pm = \omega_0 + \omega_{1\pm}$. 
The interferometer will be complete
when the two packets cross at time $t_c$, with 
$x_+(t_c) = x_-(t_c)$. 
The crossing time is approximately equal to the periods for the two packets,
but differs by small amounts
$\delta t_\pm = t_c - 2\pi/\omega_\pm$.
In terms of the frequency shifts $\omega_{1\pm}$, this implies
\be
\delta t_+ - \delta t_- = \frac{2\pi}{\omega_0^2}
\left(\omega_{1+} - \omega_{1-}\right).
\ee

Since the $\delta t$'s are already first-order in $\lambda_n$, 
their effect on the trajectories and phases can be calculated in zeroth
order. The trajectory is 
$x_\pm(t) = x_a \cos\omega_\pm t + (u_a\pm u_0)\sin\omega_\pm t$,
so setting $x_+ = x_-$ at $t_c$ yields 
$(u_a+u_0)\delta t_+ = (u_a-u_0) \delta t_-$, leading to
\be
\delta t_\pm = \frac{\pi \lambda_n h_n}{\omega_0^3}
\left(-\frac{u_a}{u_0}\pm 1\right)
\left(A_+^{n-2} - A_-^{n-2}\right).
\ee

The zero-order phase shift developed over a short time $\delta t$ 
is readily calculated from the action to be
\be
\delta\phi = -\frac{m\omega_0^2}{2\hbar} A^2 \delta t \cos 2\theta.
\label{freephase}
\ee
Here $A_\pm^2 \cos2\theta_\pm = x_a^2 - (u_a\pm u_0)^2$.
The additional phase difference from the trajectories is then 
\be
\delta\phi_+-\delta\phi_- = -\frac{\pi m \lambda_n h_n}{\hbar\omega_0}(A_+^{n-2}-A_-^{n-2})
(x_a^2+u_a^2-u_0^2)
\ee
This adds to the the result \eqref{evenphase} to give
the total anharmonic phase for a full-cycle interferometer,
\begin{align}
\Delta\phi_1^{(n)} & = \phi_{1+}^{(n)}
+ \delta\phi_+
-\phi_{1-}^{(n)} 
-\delta\phi_-
\qquad (n~\text{even})
\nonumber \\
& = \frac{\pi m \lambda_n h_n}{\hbar \omega_0} \left[
\left(1-\frac{2}{n}\right)\left(A_+^{n}-A_-^{n}\right)\right. \nonumber \\
& \hspace{5em} 
\left.-\left(x_a^2+u_a^2-u_0^2\right)
\left(A_+^{n-2}-A_-^{n-2}\right)
\rule[-0.5ex]{0pt}{4ex}\right]
.
\label{even_phase}
\end{align}
We note that in practice, the determination of
$t_c$ is subject to experimental
uncertainty, which can lead to additional phase shifts.
This issue is discussed further in Section \ref{wuncert} below.

From Eq.~\eqref{even_phase}, the first-order anharmonic effect can be 
calculated for any power $n$. In particular, it is evident that a nonzero
phase shift can be obtained only when $v_a = \omega_0 u_a$ is nonzero,
since otherwise the amplitudes $A_+$ and $A_-$ will be equal and both
packets will trace out the same orbit through the trap.
The same argument can be made for the more general result
of Eq.~\eqref{phase2}.

The values of $n$ most likely to be of interest
are  4 and 6. Evaluating \eqref{even_phase} for these cases
yields
\be
\Delta\phi_1^{(4)} = 3\pi \frac{m\lambda_4}{\hbar \omega_0} u_0^3 u_a
\label{phi4}
\ee
and
\be
\Delta\phi_1^{(6)} = 5\pi \frac{m\lambda_6}{\hbar \omega_0} u_0^3 u_a
(x_a^2+u_a^2+u_0^2).
 \label{hex}
\ee

\begin{figure}
\includegraphics[width=3.5in]{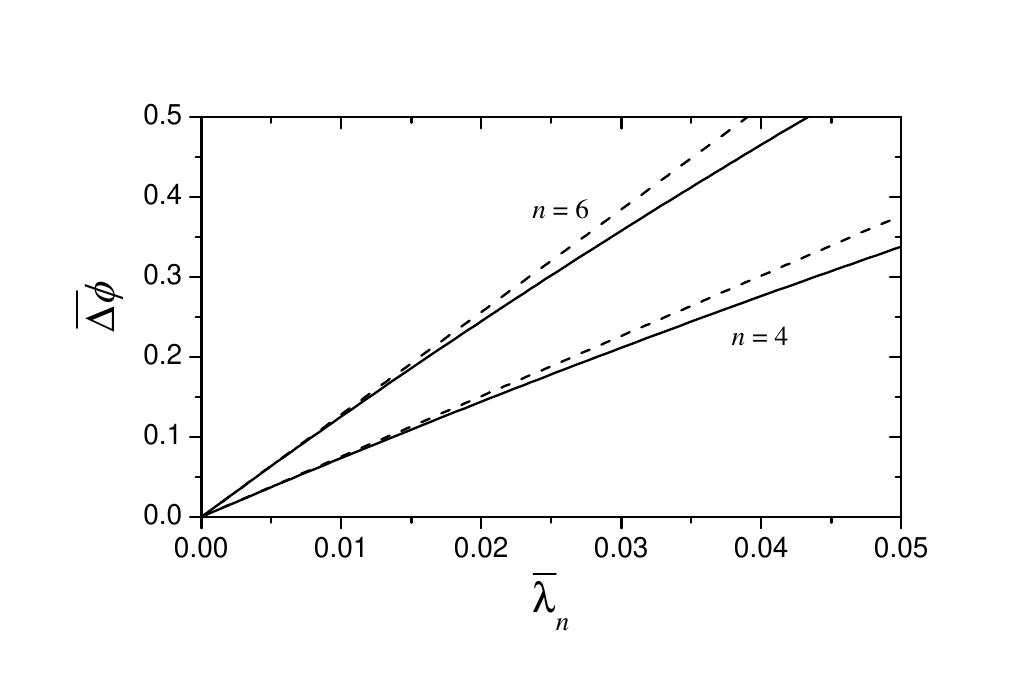}
\caption{
\label{even}
Comparison of first-order analysis (dashed curves)
to numerical calculation (solid curves) of the
anharmonic phase shifts in a free-oscillation interferometer. 
Here $\overline{\Delta\phi} = (\omega_0/\omega_r)\Delta\phi$,
with $\Delta\phi$ from \protect\eqref{even_phase}.
The recoil frequency for the Bragg laser
beam is $\omega_r = mv_0^2/8\hbar$. The dimensionless anharmonic
coefficient $\overline{\lambda}_n$ is given by $\lambda_n v_0^{n-2}/\omega_0^n$.
The curves are labeled with the power of the anharmonic perturbation $n$.
Both cases shown use $v_a = 0.1 v_0$ and $x_a = 0.1 v_0/\omega_0$.
}
\end{figure}

We checked these calculations
by comparing to 
numerical integration of the action \eqref{action2} using
the exact equation of motion \eqref{motion}. 
Figure~\ref{even}
shows a comparison of the numerical and analytical results. In all
cases examined, we found good agreement.

\subsection{Anharmonic Trap: Second-Order}
\label{secondorder}

As noted previously, there is no first-order phase shift for a power-law
perturbation with odd power $n$. However, if an odd-$n$ coefficient
$\lambda_n$ is relatively 
large, then 
the second-order effect from it could be comparable
to the first-order effect from an even power at higher $n$. In particular,
terms scaling with $(\lambda_3)^2$ have the same amplitude dependence as those
scaling with $\lambda_4$, so a second-order calculation of the odd-$n$
terms is needed to make a fair comparison of their impact.

To proceed, we shall require 
the order-one correction to the motion, $x_1$. Its equation of
motion is
\be
\ddot{x_1} + \omega_0^2 x_1 = -\lambda_n A^{n-1} \cos^{n-1}(\omega_0 t-\theta).
\label{x1eqn}
\ee
The cosine term here can be expanded using
\be
(\cos u)^p = \sum_s^p h^{(p)}_s \cos su,
\label{expansion}
\ee
where the sum is over odd $s$ from 1 to $p$ when $p$ is odd,
and over even $s$ from 0 to $p$ when $p$ is even. 
The expansion coefficients are
\be
h^{(p)}_s = \frac{p!}{(p+s)!!(p-s)!!} (2-\delta_{s0}).
\ee
Here $h_0^{(n)}$ is identical to $h_n$ in Eq.~\eqref{hn}.
The solution to \eqref{x1eqn} is then
\be
x_1 = \frac{\lambda_n A^{n-1}}{\omega_0^2}
\sum_{\text{even}~s}^{n-1} \frac{h_s^{(n-1)}}{s^2-1}\cos s(\omega_0t-\theta).
\label{x1}
\ee
For later use, we note that the
same result holds for even $n$, except that $\omega_0 \rightarrow \omega$ 
and the sum is over odd $s>1$.

Extending the perturbation series to second order, we have
$x = x_0 + x_1 + x_2$ with $\omega = \omega_0 + \omega_2$.
Inserting into the equation of motion \eqref{motion} and collecting second order
terms gives
\begin{align}
\ddot{x_2} + \omega_0^2 x_2 =  
-& (n-1)\lambda x_1 A^{n-2}\cos^{n-2}(\omega_0t-\theta) \nonumber \\
& +2\omega_0\omega_2A\cos(\omega_0t-\theta).
\end{align}
Again, the right-hand side must have no $\omega_0$ components. 
After some manipulation, this leads to
\be
\omega_2 = 
\frac{(n-1)\lambda_n^2 A^{2n-4}}{2\omega_0^3}
\sum_{\text{even}~s}^{n-1} \frac{[h_s^{(n-1)}]^2}{s^2-1}(1+\delta_{s0}).
\label{odd_dw}
\ee

To calculate the phase shift, we expand the Lagrangian to second order
and discard terms that trivially integrate to zero. This yields
\begin{align}
\phi_2^{(n)} = \frac{m}{\hbar} & \int_0^{\tau}   
\left[ \frac{1}{2} \left(\dot{x}_0^2 - \omega_0^2 x_0^2\right) \right. \nonumber \\
& \left. +\frac{1}{2} \left(\dot{x}_1^2 - \omega_0^2 x_1^2\right) -\lambda_n x_0^{n-1} x_1 \right]\,dt.
\end{align}
The first pair of terms are evaluated just as for even $n$ to give a
contribution
$\phi_A = \pi m A^2 \omega_2/\hbar$.
The second pair of terms can be evaluated using expression
\eqref{x1} and orthogonality to obtain
\begin{align}
\phi_B & = -\frac{\pi}{2} \frac{m\lambda^2 A^{2n-2}}{\hbar\omega_0^3}
\sum_{\text{even}~s}^{n-1} \frac{[h_s^{(n-1)}]^2}{1-s^2}(1+\delta_{s0})
\nonumber \\
& = \frac{\pi}{n-1}\frac{mA^2\omega_2}{\hbar}.
\end{align}
The final term can be evaluated similarly by expanding the
$\cos^{n-1}(\omega_0t-\theta)$ factor to yield
\be
\phi_C = -\frac{2\pi}{n-1}\frac{mA^2\omega_2}{\hbar}.
\ee
Combining these three terms gives the net phase for odd $n$,
\be
\phi_2^{(n)} = \frac{\pi(n-2)}{n-1} \frac{mA^2\omega_2}{\hbar}
\qquad (n~\text{odd}).
\label{odd_phase}
\ee
with $\omega_2$ from Eq.~\eqref{odd_dw}.

\begin{figure}
\includegraphics[width=3.5in]{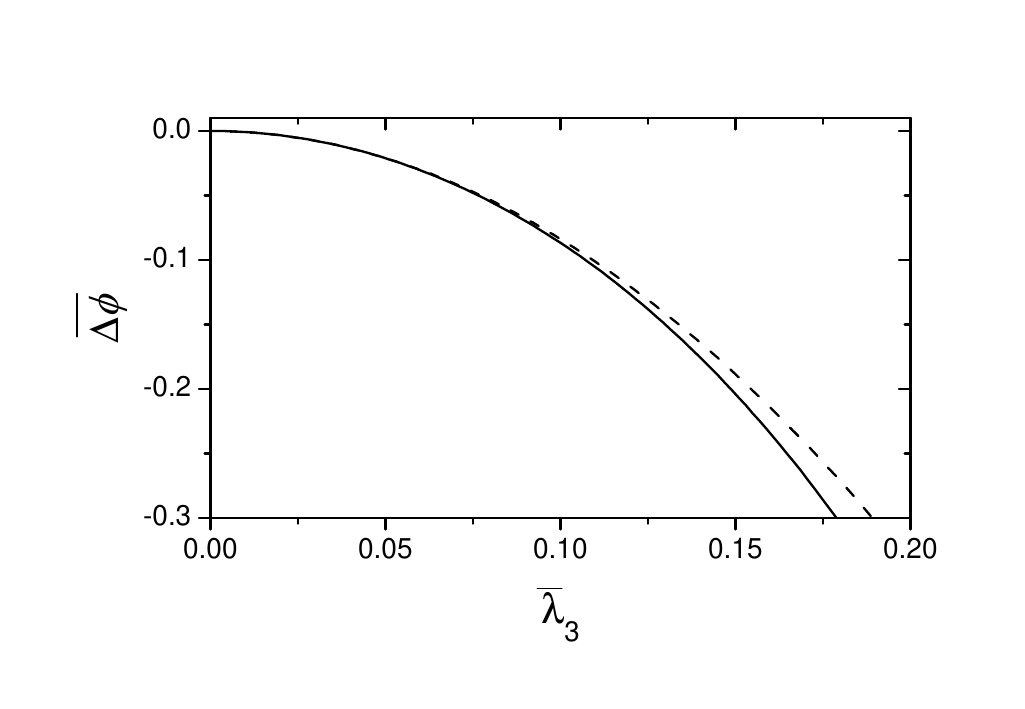}
\caption{
\label{odd}
Comparison of leading-order analysis (dashed)
to numerical calculation (solid) of the
$n = 3$ anharmonic phase. The dimensionless quantities
$\overline{\Delta\phi}$
and $\overline{\lambda}_3$ are defined as in Fig.~\protect\ref{even},
with $\Delta\phi$ from Eq.~\protect\eqref{cubic}.
The calculation uses $v_a = 0.1 v_0$ and $x_a = 0.1 v_0/\omega_0$.
}
\end{figure}

To \eqref{odd_phase} must be added the correction $\delta\phi$ accounting
for the difference in oscillation frequencies for the two trajectories
in the interferometer.
This proceeds exactly as for even $n$. In fact, the leading-order phase
shift for any $n$ can be expressed as
\begin{align}
\Delta\phi^{(n)} 
& = \frac{\pi m}{\hbar} \left[
c_n\left(\delta\omega_+ A_+^{2}-
\delta\omega_- A_-^{2}\right)\right. \nonumber \\
& \hspace{3em}
\left.-\left(x_a^2+u_a^2-u_0^2\right)
\left(\delta\omega_+-\delta\omega_-\right)
\rule[-0.5ex]{0pt}{4ex}\right]
\end{align}
where for even $n$, $c_n = (n-2)/n$ and $\delta\omega = \omega_1$, while for
odd $n$, $c_n = (n-2)/(n-1)$ and $\delta\omega = \omega_2$. The most
interesting odd case $n = 3$ gives
\be
\Delta\phi_2^{(3)} = -\frac{10\pi}{3} \frac{m\lambda_3^2}{\hbar \omega_0^3}
u_0^3 u_a,
\label{cubic}
\ee
which is similar to Eq.~\eqref{phi4} as claimed. We again checked this result
against direct numerical calculation of the phase and found
good agreement, as seen in Fig.~\ref{odd}.

\begin{figure}
\includegraphics[width=3.5in]{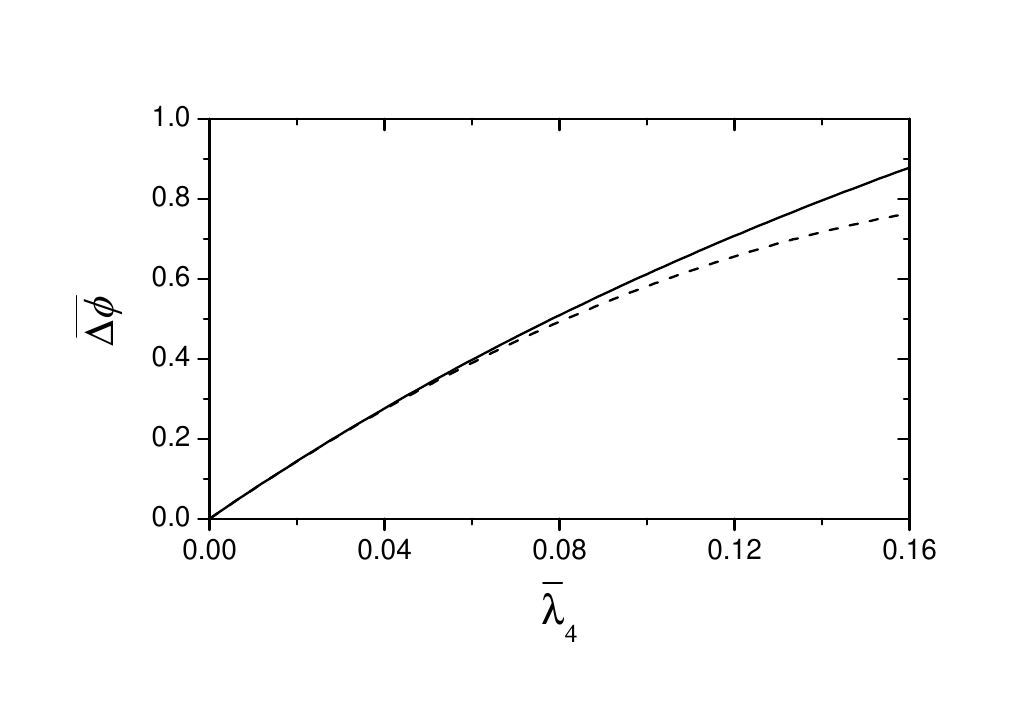}
\caption{
\label{next}
Comparison of second-order fit (dashed curves)
to numerical calculation (solid curves) 
for a full-cycle interferometer with quartic ($n = 4$) anharmonicity.
Plotted values are scaled as in Fig.~\protect\ref{even},
with $\Delta\phi$ from Eqs.~\protect\eqref{phi4} and
\protect\eqref{nlo} using $v_a = 0.1 v_0$ 
and $x_a = 0.1 v_0/\omega_0$.
}
\end{figure}

In principle, the analytical approach used here could be continued to
provide the next-leading-order correction to the phase. However,
the calculation rapidly becomes complicated \cite{Eminhizer76,Kuznetsov05}. 
We did, however, numerically
investigate the second-order correction for the quartic potential, 
as this seems to be the case most likely to have practical importance.
We find a net interferometer phase of
\be
\Delta\phi_2^{(4)} = 
-K\frac{m\lambda_4^2}{\hbar\omega_0^3} u_0^3 u_a 
\left(u_0^2+\kappa_1 u_a^2 + \kappa_2 x_a^2 + \kappa_3 x_a u_a\right)
\label{nlo}
\ee
with dimensionless $K = 20.61$, $\kappa_1 = 1.667$, 
$\kappa_2 = 1.001$, and $\kappa_3 = 2.157$. The estimated accuracy 
for $\Delta\phi_2$ is 0.2\%, based on the 
consistency of the fit across a range of parameter values.
Figure~\ref{next} shows an example comparison including both first and second
order contributions. We note that
\eqref{hex} and \eqref{nlo} have similar scaling with amplitude,
so if neither $\lambda_4$ nor $\lambda_6$ are suppressed, then both
terms must be considered if this level of accuracy is required.

\subsection{Half-Cycle Interferometer}

Until now we have supposed the atoms to complete a (nearly) full
oscillation in the trap. An interferometer can also be implemented
using just one half oscillation, with the packets recombined at
$t \approx \tau/2$.
Here the phase will clearly be more sensitive to 
asymmetries in the confining potential, but it also becomes sensitive to
asymmetric effects that might be of interest. It is worth noting,
however, that a half-cycle interferometer remains insensitive to
uniform static forces such as gravity, since a constant force will
simply shift the center of the harmonic potential with no effect on the
trajectories or phase.

For even $n$, the phase calculation proceeds just
as in the full-oscillation case, but all terms are reduced in magnitude
by a factor of two. The final phase shift is therefore just one
half of Eq.~\eqref{even_phase}.

The effect of odd anharmonicities changes more significantly, as 
expected. There is now a first-order 
contribution to the phase,
\be
\phi\sub{half}^{(n)} = -\frac{m}{\hbar}
\int_0^{\tau/2} \frac{\lambda_n}{n} x_0^n \,dt
\ee
that is readily evaluated to 
\be
\phi\sub{half}^{(n)} = -\frac{2m\lambda_n}{n\hbar\omega_0}h_1^{(n)} A^n\sin\theta.
\ee
To this order, there is no correction to the oscillation frequency,
so $\phi\sub{half}^{(n)}$ is the only contribution to the interferometer
phase difference. Here 
$A_\pm \sin\theta_\pm = u_a \pm u_0$. Evaluating $\phi\sub{half}$
for the case $n = 3$ yields
\be
\Delta\phi\sub{half}^{(3)} = -\frac{2m\lambda_3}{\hbar\omega_0} u_0
\left(u_0^2 + x_a^2 + 3u_a^2\right).
\ee
Here the lack of symmetry means that a non-zero effect is
obtained even for $u_a = 0$.

\section{Discussion and Impact}
\label{impact}

\subsection{Estimation of Effect}

Evaluating the anharmonic phase shift for a particular experiment
obviously requires knowing the value of the lowest-power unsuppressed 
$\lambda_n$ coefficient, either by measurement or by calculation from the
trap geometry. 
Typically, however, an order-of-magnitude estimate for $\lambda_n$ can
be obtained from the power-series expansion of Eq.~\eqref{power}.
For the appropriate length scale $R$, the coefficients $f_n$
in that expansion should have comparable magnitudes.
Furthermore, in the typical case that the harmonic confinement 
is provided by the same elements that introduce $f(x)$, the
expansion could be extended to $n = 2$ and the $f_2$ coefficient,
given by $f_2 = \omega_0^2 R^2/2$, will have comparable magnitude
to the other $f_n$'s. In terms of the $\lambda_n$ coefficients of
Eq.~\eqref{lambda}, this implies
\be
\lambda_n \approx  \frac{n\omega_0^2}{2R^{n-2}}.
\label{approx}
\ee

This allows an estimation as to when anharmonic
effects are likely to be important. For instance, a
symmetric trap (with $\lambda\sub{odd} \approx 0$) would have
\be
\Delta\phi_1^{(4)} \approx 6\pi \frac{m v_a v_0^3}{\hbar \omega_0^3 R^2}
\ee
Using Rb atoms with $\omega_0 = 2\pi\times 10$~Hz, $v_0 = 1.2$~cm/s,
and $R = 1$~cm, maintaining $\Delta\phi \ll 1$ requires
the initial atomic velocity $v_a \ll 0.5$~mm/s. The same configuration
with $\omega_0$ reduced to $2\pi\times 1$~Hz would require
$v_a \ll 0.5~\mu$m/s. The strong dependence on $\omega_0$ reflects the fact
that a weaker trap will allow the atomic trajectories to
extend to larger distances where the anharmonicity is more significant.

In principle, a fixed phase shift could be measured and subtracted out,
but in practice $v_a$ is likely to fluctuate, making the anharmonicity
into a source of noise.
For thermal atoms, the velocity spread will be determined by the gas 
temperature, while for condensate atoms it is limited by the uncertainty
principle and technical effects.
In our experiments \cite{Reeves05}, 
we produce Rb condensates in a relatively tight trap
and then adiabatically reduce the confinement to give an oscillation
frequency in the range of 1 to 10 Hz. After this process, we 
typically observe a center-of-mass motional excitation
corresponding to a velocity variance
$\sigma_v^2$ of about $\omega_0 \times 10^{-8}~\text{m}^2/\text{s}$. 
We attribute this to a combination of imperfect adiabaticity, 
forces from uncontrolled ambient magnetic fields, and mechanical vibrations
of the apparatus.
In comparison, the non-interacting harmonic oscillator ground state
has $\sigma_v^2/\omega_0 = \hbar/m = 7.3\times 10^{-10}~\text{m}^2/\text{s}$.
The fundamental 
velocity uncertainty can be even lower in an interacting condensate
\cite{Stenger99},
but the interplay between anharmonicity and interactions requires
additional consideration beyond the scope of this work.

In practice, then, anharmonic effects can be expected to limit the usable
oscillation frequency for a given trap geometry, and thus 
the measurement time $\tau$ of the interferometer. In the
case of quartic anharmonicity with our empirical
$\sigma_v$, the phase fluctuations will reach one radian at
$\tau \approx (0.09~\text{s})R^{4/5}$, for $R$ in cm.
This increases by about a factor of two for the
ideal ground state $\sigma_v$.
Anharmonic effects can thus be expected to 
impact a cm-scale device 
operating with interaction times greater than about 0.1~s.

\subsection{Amelioration}

An obvious way to reduce the impact of trap anharmonicity is to
reduce the anharmonicity itself by using a larger trap geometry
to increase $R$. However, practical applications often favor
a more compact apparatus. Also, in both magnetic and optical traps the
electrical or optical power required increases rapidly with the trap size.

Alternatively, a small trap could be designed with reduced anharmonicity.
Typically this would be achieved by tuning one or more $\lambda_n$
coefficients to zero. 
For example, 
with our empirical $\sigma_v$ 
a trap with $\lambda_4 \approx 0$ and
$\lambda_6 \approx 3\omega_0^2/R^4$ would give
one radian of phase noise at 
$\tau \approx (0.5~\text{s})R^{8/9}$, again with $R$ in cm.
This is about five times better than that 
obtained with unsuppressed $\lambda_4$.
Controlling many coefficients in this manner, however, 
is likely to be challenging.

\begin{figure}
\includegraphics[width=3.5in]{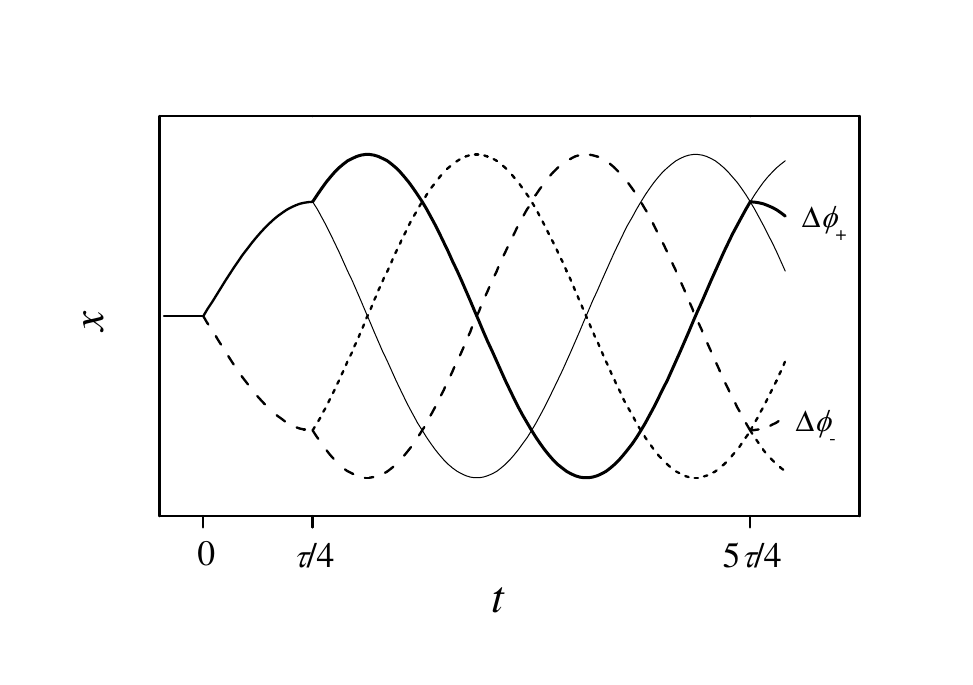}
\caption{
\label{dual}
Dual interferometer configuration. At time $t = 0$, a Bragg splitting pulse is
applied to the nominally stationary initial wave packet in a nominally harmonic
trap of period $\tau$. The resulting two packets move in the potential until
they come to nominal rest at time $\tau/4$, when the splitting pulse is again
applied. This produces two pairs of packets.
All four packets are 
allowed to propagate for a measurement time $\tau$ and then recombined
with a final pulse. 
The upper pair of trajectories
(solid curves) forms one interferometer with output phase
$\Delta\phi_+$, and the lower pair (dashed curves)
forms another with output phase $\Delta\phi_-$. The phase difference
$\Psi = \Delta\phi_+-\Delta\phi_-$ features reduced sensitivity to
the anharmonicity of the trap potential.
}
\end{figure}

We propose here another
possibility for controlling anharmonic effects: the use of
a dual interferometer, as illustrated in Fig.~\ref{dual}.
Here a wave packet initially at $(x_a, v_a)$ is split and allowed
to propagate for one quarter oscillation,
after which the packets will be
nearly at rest, with residual velocities $v \approx -\omega_0 x_a$
for both. These packets are 
used as the sources for two independent interferometers:
the packets are split, allowed to
oscillate for one period, and then recombined.
The output signal $\Psi$ is taken as the difference between the 
phases of two interferometers.
Because the initial velocities of the two interferometers are
correlated, the leading-order anharmonic phases cancel in the difference.
The phase difference is also less sensitive to mechanical vibrations
and several other technical effects \cite{Foster02}.
However, this 
configuration evidently requires the phase shift of metrological 
interest to be differential between the two interferometers.
For example, a local field might be applied to just a single packet.

We analyze the dual interferometer for the $n=4$ case. The
trajectories are, from Eq.~\eqref{x1},
\be
x(t) = A\cos(\omega t-\theta)
+\frac{\lambda_4}{32\omega_0^2} A^3\cos3(\omega t-\theta).
\label{traj2}
\ee
We use this solution to determine the actual initial conditions for 
the subsequent pair of interferometers. We here use $+$ $(-)$ to label the
interferometer derived from the packet originally given a positive
(negative) momentum kick.

The amplitudes and angles in \eqref{traj2} must be determined
in terms of $x_\pm(0) = x_a$ and $u_\pm(0) = u_a \pm u_0$.
We write
$A = A_0+A_1$ and $\theta = \theta_0+\theta_1$.
In zeroth order, we have
$x_a = A_{0\pm}\cos\theta_{0\pm}$ and
$u_a \pm u_0 = A_{0\pm}\sin\theta_{0\pm}$. The
first-order corrections are then calculated to be
\be
A_{1\pm} = \frac{1}{32}\frac{\lambda_4 A_{0\pm}^3}{\omega^2_0}
\left(8\cos^4\theta_{0\pm}-9 \right).
\ee
and
\be
\theta_{1\pm} = -\frac{1}{8}\frac{\lambda_4}{\omega_0^2} 
A_{0\pm}^2\cos\theta_{0\pm}\sin\theta_{0\pm}
\left(3+2\cos^2\theta_{0\pm} \right).
\ee
The frequencies $\omega_\pm$ are given by \eqref{w1_even},
\be
\omega_\pm  = \omega_0+\frac{3\lambda_4}{8\omega_0}A_{0\pm}^2.
\ee

After the first splitting pulse, the packets propagate
for a time $t_b = \pi/(2\bar{\omega})$, where 
$\bar{\omega} = (\omega_++\omega_-)/2$.
Inserting this into the above trajectories yields
\begin{align}
u_{\pm}(t_b) = -x_a - \frac{\lambda_4}{8\omega_0^2}
& \left[ 3\pi u_a u_0^{\rule{0pt}{1ex}}\left(u_0\pm u_a\right) \right. \\
& \left. + 2x_a\left(3u_0^2 + 3u_a^2 + x_a^2 \pm 6u_0 u_a\right)\right]. 
\nonumber
\end{align}
Using these as the initial velocities for the two subsequent interferometers
in \eqref{phi4} gives
a phase difference
\be
\Psi \equiv \Delta\phi_+ - \Delta\phi_- = -\frac{9\pi}{4} 
\frac{m\lambda_4^2}{\hbar\omega_0^3}u_0^4 u_a (\pi u_a + 4x_a),
\ee
which is much smaller than the individual $\Delta\phi$'s.
The result is second-order in $\lambda_4$, but the direct
second-order effect of \eqref{nlo} also largely cancels 
to give a third-order correction to $\Psi$.
The leading-order $\lambda_6$ effect \eqref{hex} cancels
as well.
Using the estimated value $\lambda_4 \approx 2\omega_0^2/R^2$,
our empirical values for $\sigma_v$, and taking 
$\sigma_x \approx \sigma_v/\omega_0$,
we obtain a measurement time limit 
$\tau \approx 0.9 R$~s/cm for $\sigma_\Psi < 1$. 

\subsection{Measurement of Anharmonicity}
\label{wuncert}

Although the estimates above may be helpful, an accurate consideration
of the effects discussed here will require actual knowledge
of the relevant anharmonic coefficients. While they can be calculated
in principle, an experimental
technique to measure them would likely be useful.
The most straightforward approach is to 
observe how the oscillation frequency depends on the motional amplitude,
via $\delta\omega$. However, 
it may be difficult to measure $\delta\omega$ with sufficient accuracy.
For example, in a 10-Hz trap with quartic anharmonicity
and $R = 1$~cm ($\lambda_4 = 8\times 10^3$~cm$^{-2}$~s$^{-2}$),
Rb atoms with 
$v_a = 0.5$~mm/s would experience a significant
phase $\Delta\phi \approx 1$~rad.
However, the frequency shift for atoms with amplitude 
$A = v_0/\omega_0 \approx 200~\mu$m would
be only 3 mHz. 
Such a small shift could be difficult to measure, given a
finite lifetime of atoms in the trap.
It may prove more effective to use the interferometer
to characterize the trap, by for instance measuring how
the interference phase $\Delta\phi$ varies with $v_a$.
The results presented here should be useful for this purpose as well.

It should also be noted that imprecise knowledge of the oscillation
frequency $\omega$ can itself lead to phase uncertainty. In
deriving \eqref{even_phase}, we assumed a measurement time $t_c$ such
that the two packet centers exactly crossed at the time of recombination. 
If the actual measurement time is too far from $t_c$, then the interference
contrast will be reduced because the packets will not be well-overlapped. For
a small timing error, however, the overlap will remain large and
the dominant effect will be a phase
shift resulting from the differing velocities of the two packets.
This can be calculated from Eq.~\eqref{freephase} along with a correction
$mv_a \delta x/\hbar$ for packet separation $\delta x$. The result is
\be
\delta\phi = 2\frac{m}{\hbar} v_0 v_a \delta t
\ee
for timing error $\delta t = t-t_c$.
Using our empirical value for the velocity uncertainty in Rb,
this results in a significant phase uncertainty of 
$(7~\text{rad/ms})\delta t$.
As above, determining $\omega$ with sufficient accuracy to avoid this problem
may be challenging. This phase error can be reduced using
the dual interferometer scheme, because to lowest order it is 
the same for both pairs of trajectories.

\section{Conclusions}
\label{conclusion}

We have calculated the leading-order effects of trap anharmonicity
on a free-oscillation atom interferometer. For a typical cm-scale
device, anharmonic phase shifts are likely to be important for interaction
times of about 0.1~s or greater, with the effects growing rapidly as
the interaction time is increased.
Possible methods for amelioration include
nulling the 
low-order anharmonic coefficients via careful trap design,
using a larger-scale trap with less anharmonicity, minimization of
the initial velocity of the atoms, or phase cancellation in a dual 
interferometer. 

\begin{acknowledgments}
This work was supported by the National Science Foundation (Grant No. PHY-
0244871).

\end{acknowledgments}


\end{document}